# PARAMETRIC SYNTHESIS OF QUANTUM CIRCUITS FOR TRAINING PERCEPTRON NEURAL NETWORKS

**Cesar Borisovich Pronin**

**Andrey Vladimirovich Ostroukh**

MOSCOW AUTOMOBILE AND ROAD CONSTRUCTION STATE TECHNICAL UNIVERSITY (MADI)., 64, Leningradsky prospect, Moscow, Russia

**Abstract:** This paper showcases a method of parametric synthesis of quantum circuits for training perceptron neural networks. Synapse weights are found using Grover's algorithm with a modified oracle function. The results of running these parametrically synthesized circuits for training perceptrons of three different topologies are described. The circuits were run on a 100-qubit IBM quantum simulator.

The synthesis of quantum circuits is carried out using quantum synthesizer "Naginata", which was developed in the scope of this work, the source code of which is published and further documented on GitHub [1]. The article describes the quantum circuit synthesis algorithm for training single-layer perceptrons.

At the moment, quantum circuits are created mainly by manually placing logic elements on lines that symbolize quantum bits.

The purpose of creating Quantum Circuit Synthesizer "Naginata" was due to the fact that even with a slight increase in the number of operations in a quantum algorithm, leads to the significant increase in size of the corresponding quantum circuit. This causes serious difficulties both in creating and debugging these quantum circuits.

The purpose of our quantum synthesizer is enabling users an opportunity to implement quantum algorithms using higher-level commands. This is achieved by creating generic blocks for frequently used operations such as: the adder, multiplier, digital comparator (comparison operator), etc. Thus, the user could implement a quantum algorithm by using these generic blocks, and the quantum synthesizer would create a suitable circuit for this algorithm, in a format that is supported by the chosen quantum computation environment. This approach greatly simplifies the processes of development and debugging a quantum algorithm.

## Introduction

The goal of project "Naginata - Quantum Circuit Synthesizer" [1] is to create a prototype system for synthesizing complex quantum circuits. One of the applications of this system will be in parametric synthesis of quantum circuits for finding the weights of a neural network (perceptron) of a given topology. Quantum circuits synthesized by the program are exported to a *.qasm text file in the form of OPENQASM 2.0 code, compatible with IBM Quantum - a cloud quantum computing environment.

This paper describes the results of running synthesized circuits for training perceptrons of three different topologies, which were set as parameters in the file "qnn_generation_tests.py". The circuits were tested on a 100-qubit IBM "simulator_mps" quantum simulator in the IBM Quantum cloud environment.

The circuit for finding the weights of a perceptron is synthesized based on Grover's algorithm [3], for which a custom oracle is created based on given parameters of the perceptron's topology and input neuron values.

## Algorithm of synthesizing quantum circuits for training a single layer perceptron

The topology of the required neural network is inputted as two lists of dictionaries (python). The first list describes connections between input and hidden neurons. The second list describes connections between hidden and output neurons.

The "build_param_network" function creates all necessary quantum registers to store the desired weights, and also independently combines them into "grover_regs" - a list of registers that store the "body" of the Grover algorithm. Then, registers are created and filled to store the values of input neurons and the target value Ac. Next, registers are created to store multiplication results, from them, some get selected to hold the total sum of these results for each hidden and for each output neuron.

The first step in constructing the circuit is applying the Hadamard gates to the weight registers $w_n$ (the first step in Grover's algorithm).

The next step, is the construction of the oracle function based on the topology of the neural network:

1) The oracle function contains calculations (Mul - operation of multiplication, Sum - the operation of addition) that are performed during the forward pass of the neural network - calculating the values of hidden and then output neurons (1). With a linear activation function of hidden neurons $H_{x\ input} = H_{x\ output} = H_x$ .

$$H_1 = I_1 w_1 + I_2 w_2$$
$$H_2 = I_3 w_3 + I_4 w_4 \qquad (1)$$
$$O_1 = H_1 w_5 + H_2 w_6$$

2) The values of the output neurons are compared with the threshold value Ac, which is a condition for selecting suitable values $w_n$.

3) The registers that store the results of comparison operations are rotated by a NCZ gate. N-Controlled-Z Gate (NCZ) – a Pauli-Z gate, with multiple (N) control qubits.
4) Reversed versions of the calculation operations from steps 1 and 2 are performed

The last step is to perform the amplitude amplification function (the third step in Grover's algorithm) on $w_n$ registers, and to measure the values of these registers. Thus, the "body" of Grover's algorithm is placed in registers $w_n$.

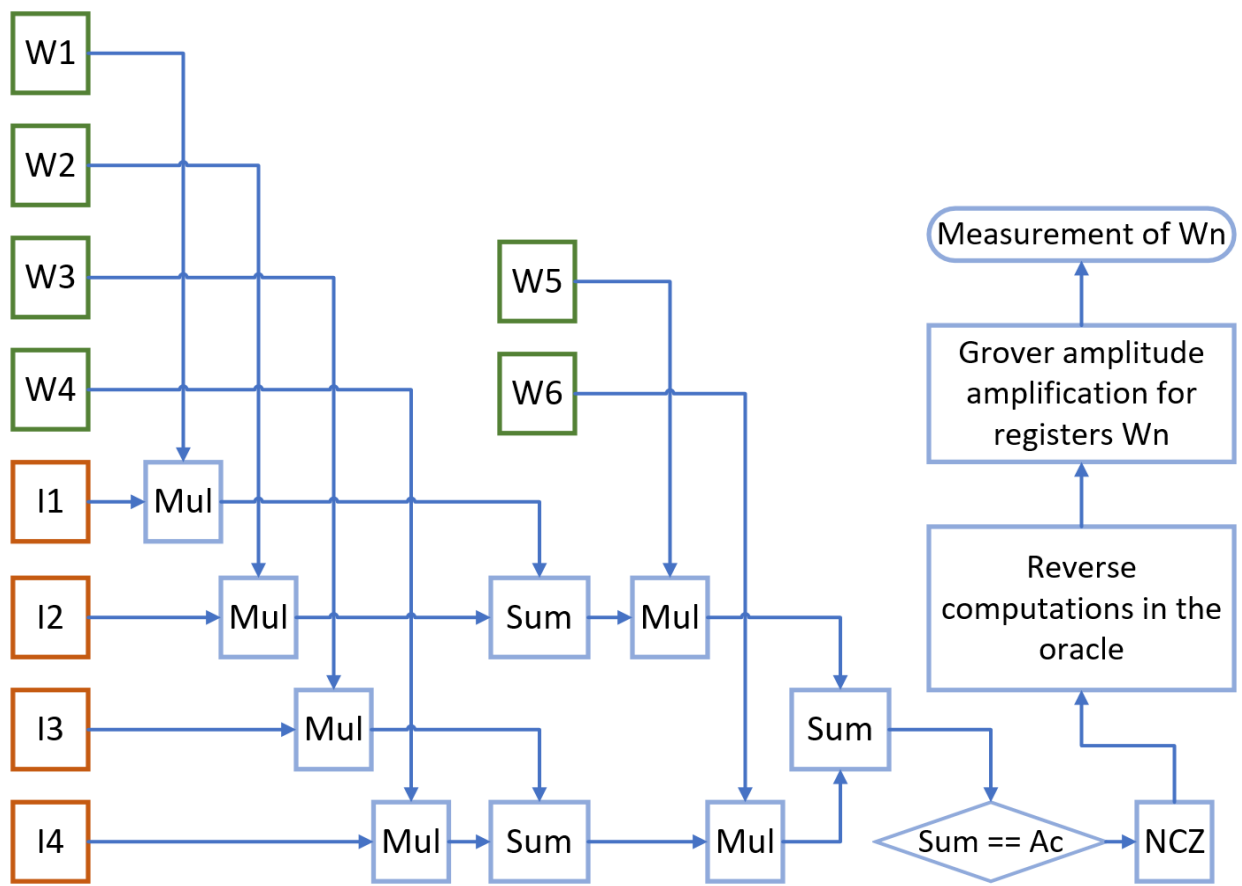

*Fig. 1.1. Diagram for calculating weights $w_n$ with Grover's algorithm*

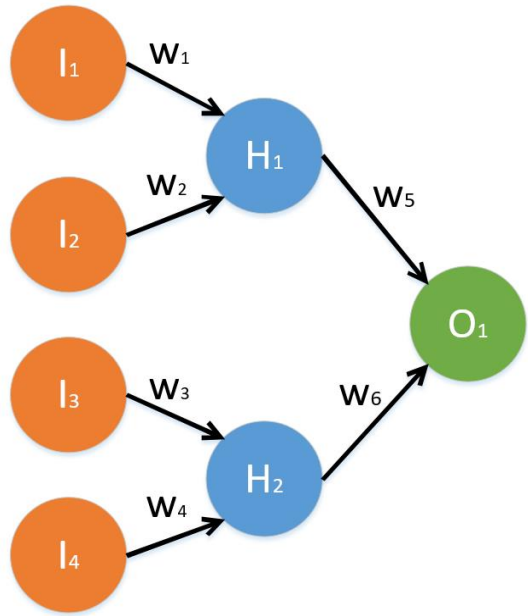

*Fig. 1.2. Perceptron topology for diagram on **Fig 1.1***

For a given neural network (**Fig. 1.2**), the condition for selecting suitable values $w_n$ is formed, as inequality (2) [4-6]:

$$(I_1 w_1 + I_2 w_2) * w_5 + (I_3 w_3 + I_4 w_4) * w_6 \geq Ac \qquad (2)$$

$$\text{or } O_1 \geq Ac$$

To simplify the analysis of resulting values, the examples select only pairs of weights $w_n$, with which $O_1 = Ac$.

**Generating a quantum circuit for training perceptron example №1**

Let's observe an example of finding weights of a perceptron with the topology shown on **Fig. 2.1**. The goal of training this neural network is to find the coefficients $w_i$ that satisfy inequality (3).

$$(I_1 w_1 + I_2 w_2) * w_3 \geq Ac \qquad (3)$$

To simplify the analysis of the algorithm's results, we implement only condition (4).

$$(I_1 w_1 + I_2 w_2) * w_3 = Ac \qquad (4)$$

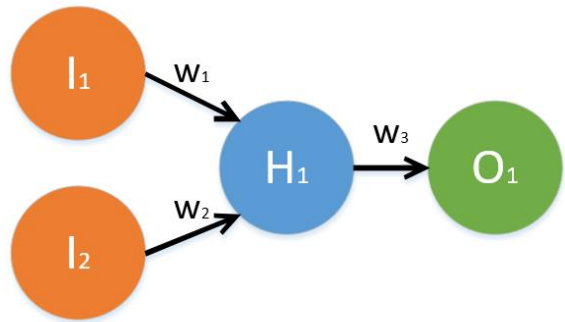

*Fig. 2.1. Topology of example perceptron №1*

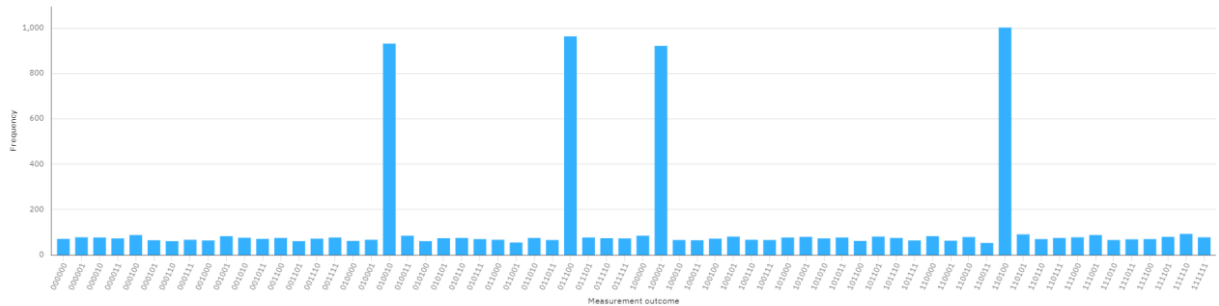

*Fig. 2.2. The spread of measured values of the algorithm for training perceptron №1 after 8192 iterations (shots = 8192)*

At the input of the algorithm, the values are: $I_1 = 11_2 = 3_{10}$; $I_2 = 10_2 = 2_{10}$; $Ac = 000110_2 = 6_{10}$

At the output of the algorithm, the following values were obtained as solutions: 010010, 011100, 100001, 110100. They correspond to peaks in the diagram on **Fig. 2.2**. Based on the order in which the registers are defined for this circuit in file "qnn_generation_tests.py" and the bit numbering order in the IBM

Quantum environment, the obtained values should be divided into equal 2-bit registers corresponding to desired values of $w_1$, $w_2$ и $w_3$, in the manner indicated in **table 1**.

Table 1. Order of splitting the measured bit strings into w values, for perceptron №1

| *measured* | $w_3$ | $w_2$ | $w_1$ |
|---|---|---|---|
| 010010 | 01 | 00 | 10 |
| 011100 | 01 | 11 | 00 |
| 100001 | 10 | 00 | 01 |
| 110100 | 11 | 01 | 00 |

**Generating a quantum circuit for training perceptron example №2**

The goal of training the neural network on **Fig. 3.1** is to find the coefficients $w_i$ that satisfy inequality (5).

$$I_1 w_1 w_3 + I_1 w_2 w_4 \geq Ac \qquad (5)$$

To simplify the analysis of the algorithm's results, we implement only condition (6).

$$I_1 w_1 w_3 + I_1 w_2 w_4 = Ac \qquad (6)$$

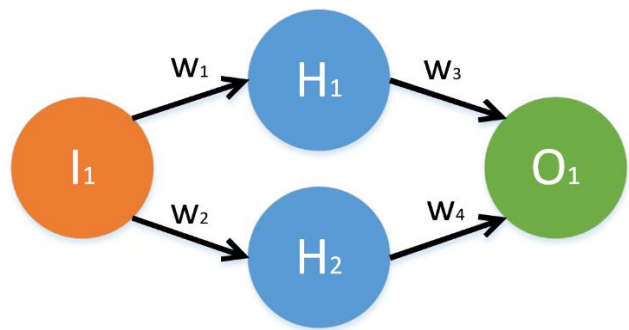

*Fig. 3.1. Topology of example perceptron №2*

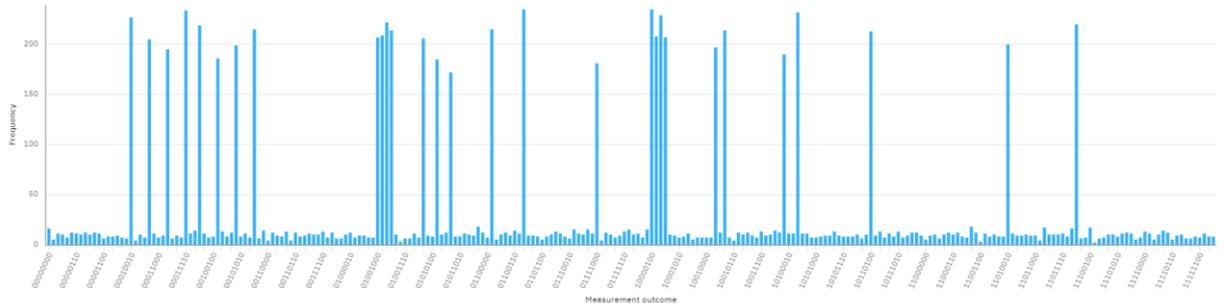

*Fig. 3.2. The spread of measured values of the algorithm for training perceptron №2 after 8192 iterations (shots = 8192)*

At the input of the algorithm, the values are: $I_1 = 11_2 = 3_{10}$; $Ac = 000110_2 = 6_{10}$

At the output of the algorithm, the values obtained after measurement are given in **table 2**. They correspond to peaks in the diagram on **Fig. 3.2**. Based on the order in which the registers are defined for this circuit in file "qnn_generation_tests.py" and the bit numbering order in the IBM Quantum environment, the obtained values should be divided into equal 2-bit registers corresponding to desired values of $w_1$, $w_2$, $w_3$ and $w_4$, in the manner indicated in **table 2**.

Table 2. Order of splitting the measured bit strings into w values, for perceptron №2

| *measured* | $w_4$ | $w_3$ | $w_2$ | $w_1$ |
|---|---|---|---|---|
| 00010010 | 00 | 01 | 00 | 10 |
| 00010110 | 00 | 01 | 01 | 10 |
| 00011010 | 00 | 01 | 10 | 10 |
| 00011110 | 00 | 01 | 11 | 10 |
| 00100001 | 00 | 10 | 00 | 01 |
| 00100101 | 00 | 10 | 01 | 01 |
| 00101001 | 00 | 10 | 10 | 01 |
| 00101101 | 00 | 10 | 11 | 01 |
| 01001000 | 01 | 00 | 10 | 00 |
| 01001001 | 01 | 00 | 10 | 01 |
| 01001010 | 01 | 00 | 10 | 10 |

| 01001011 | 01 | 00 | 10 | 11 |
| --- | --- | --- | --- | --- |
| 01010010 | 01 | 01 | 00 | 10 |
| 01010101 | 01 | 01 | 01 | 01 |
| 01011000 | 01 | 01 | 10 | 00 |
| 01100001 | 01 | 10 | 00 | 01 |
| 01101000 | 01 | 10 | 10 | 00 |
| 01111000 | 01 | 11 | 10 | 00 |
| 10000100 | 10 | 00 | 01 | 00 |
| 10000101 | 10 | 00 | 01 | 01 |
| 10000110 | 10 | 00 | 01 | 10 |
| 10000111 | 10 | 00 | 01 | 11 |
| 10010010 | 10 | 01 | 00 | 10 |
| 10010100 | 10 | 01 | 01 | 00 |
| 10100001 | 10 | 10 | 00 | 01 |
| 10100100 | 10 | 10 | 01 | 00 |
| 10110100 | 10 | 11 | 01 | 00 |
| 11010010 | 11 | 01 | 00 | 10 |
| 11100001 | 11 | 10 | 00 | 01 |

**Generating a quantum circuit for training perceptron example №3**

The goal of training the neural network on **Fig. 4.1** is to find the coefficients $w_i$ that satisfy inequality (7).

$$I_1 w_1 w_3 + I_2 w_2 w_4 \geq Ac \qquad (7)$$

To simplify the analysis of the algorithm's results, we implement only condition (8).

$$I_1 w_1 w_3 + I_2 w_2 w_4 = Ac \qquad (8)$$

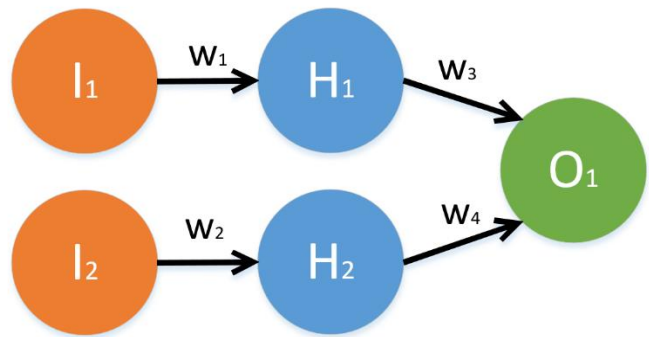

*Fig. 4.1. Topology of example perceptron №3*

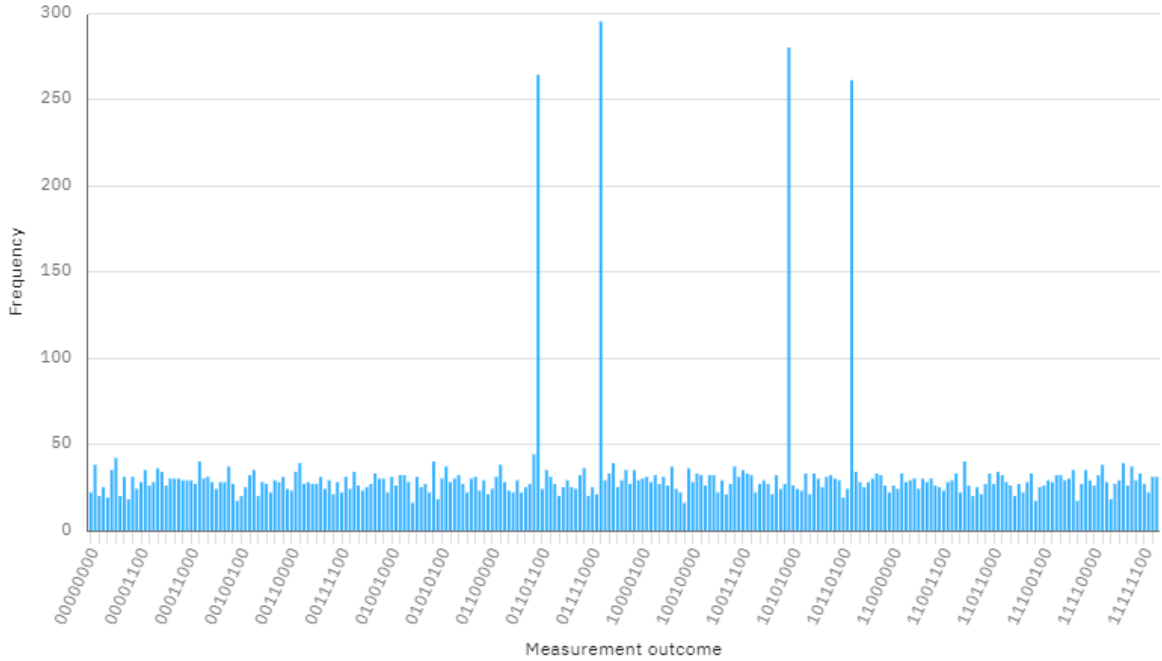

*Fig. 4.2. The spread of measured values of the algorithm for training perceptron №3 after 8192 iterations (shots = 8192)*

At the input of the algorithm, the values are: $I_1 = 11_2 = 3_{10}$; $I_2 = 10_2 = 2_{10}$; $Ac = 010110_2 = 22_{10}$

At the output of the algorithm, the values obtained after measurement are given in **table 3**. They correspond to peaks in the diagram on **Fig. 4.2**. Based on the order in which the registers are defined for this circuit in file "qnn_generation_tests.py" and the bit numbering order in the IBM Quantum environment, the obtained values should be divided into equal 2-bit registers corresponding to desired values of $w_1$, $w_2$, $w_3$ and $w_4$, in the manner indicated in **table 3**.

Table 3. Order of splitting the measured bit strings into w values, for perceptron №3

| measured | $w_4$ | $w_3$ | $w_2$ | $w_1$ |
|---|---|---|---|---|
| 01101011 | 01 | 10 | 10 | 11 |
| 01111010 | 01 | 11 | 10 | 10 |
| 10100111 | 10 | 10 | 01 | 11 |
| 10110110 | 10 | 11 | 01 | 10 |

## Conclusions

This article describes the results of running synthesized circuits for training perceptrons of three different topologies, which were specified as parameters in file "qnn_generation_tests.py" of the "Naginata" Quantum Circuit Synthesizer. The circuits were run on a 100-qubit IBM "simulator_mps" quantum simulator in the IBM Quantum cloud environment.

The circuits for finding the perceptron's weights were synthesized based on the principles of Grover's algorithm [3], and by forming specialized oracles, based on the given parameters.

The development of systems for synthesizing quantum algorithms can become one of the main directions for development and further implementation of quantum computing in various economic sectors, because they help to simplify the process of creating and debugging complex quantum circuits.

The use of computational advantages of quantum computers in machine learning could significantly optimize artificial intelligence models and improve the accuracy of their training, which in turn will make these models more reliable and versatile.

**Author details**

**Andrey Vladimirovich Ostroukh**, Russian Federation, full member RAE, Doctor of Technical Sciences, Professor, Department «Automated Control Systems». State Technical University – MADI, 125319, Russian Federation, Moscow, Leningradsky prospekt, 64. Tel.: +7 (499) 151-64-12. http://www.madi.ru , email: ostroukh@mail.ru , ORCID: https://orcid.org/0000-0002-8887-6132

**Cesar Borisovich Pronin**, Russian Federation, PhD student, Department «Automated Control Systems». State Technical University – MADI, 125319, Russian Federation, Moscow, Leningradsky prospekt, 64. Tel.: +7 (499) 151-64-12. http://www.madi.ru , email: caesarpr12@gmail.com , ORCID: https://orcid.org/0000-0002-9994-1032